\documentclass[aps,pre,showpacs,superscriptaddress,eqsecnum]{revtex4-1}

\usepackage{graphicx}
\usepackage{amsmath}
\usepackage{amssymb}

\usepackage{combelow}
\usepackage{bm}
\usepackage{slashed}
\usepackage{braket}
\usepackage{color}

\def\eqref#1{(\ref{#1})}

\def\feq{\ensuremath{f^{\rm{(eq)}}}}
\def\hh{\mathfrak{h}}

\begin{document}

\title{Lattice Boltzmann approach to rarefied gas flows using half-range Gauss-Hermite
quadratures: Comparison to DSMC results based on ab initio potentials}

\author{Victor E. \surname{Ambru\cb{s}}}
\email[E-mail: ]{victor.ambrus@e-uvt.ro}
\affiliation{Department of Physics, West University of Timi\cb{s}oara,
	Bd.~Vasile P\^arvan 4, Timi\cb{s}oara 300223, Romania}

\author{Felix \surname{Sharipov}}
\email[E-mail: ]{sharipov@fisica.ufpr.br}
\affiliation{Departamento de F\'isica, Universidade Federal do Paran\'a, Curitiba, 81531-980
Brazil}

\author{Victor \surname{Sofonea}}
\email[E-mail: ]{sofonea@gmail.com}
\thanks{Corresponding author.}
\affiliation{Center for Fundamental and Advanced Technical Research, 
Romanian Academy, \\Bd.~Mihai Viteazul 24, Timi\cb{s}oara 300223, Romania}

\begin{abstract}
In this paper, we employ the lattice Boltzmann method
 to solve 
the Boltzmann equation with the Shakhov model for the 
collision integral
in the context of the 3D planar Couette flow.
The half-range Gauss-Hermite quadrature is used to account for the 
wall-induced discontinuity in the distribution function. The lattice 
Boltzmann simulation results are compared with direct simulation Monte 
Carlo (DSMC) results for ${}
^3{\rm He}$ and ${}^4{\rm He}$ atoms 
interacting via ab initio potentials, at various values of the rarefaction 
parameter 
$\delta$, where the temperature of the plates varies from $1\ {\rm K}$ 
up to $3000\ {\rm K}$. Good agreement is observed between the results 
obtained using the Shakhov model and the DSMC data
at large values of the rarefaction parameter. The agreement deteriorates as 
the rarefaction parameter is decreased, however we highlight that the relative 
errors in the non-diagonal component of the shear stress do not exceed 
$2.5\%$.
\end{abstract}

\maketitle

\section{INTRODUCTION}

 \vspace{-2pt}

The main difficulty in simulating steady-state rarefied channel
flows is caused by the discontinuity in the distribution function
induced by the particle-wall interaction.
A generally accepted method for the simulation of rarefied flows is
the Direct Simulation Monte Carlo (DSMC) method. Although
accurate, DSMC simulations are computationally demanding,
particularly in the hydrodynamic and transition regimes.
A convenient alternative to DSMC is the Boltzmann equation with
a suitable model (e.g., BGK \cite{bhatnagar54} or Shakhov \cite{shakhov68a}) 
for the collision term \cite{sharipov16}.
Various numerical methods have been developed to solve
such model equations, including the Discrete Velocity Method
(DVM) \cite{sharipov16}, the Discrete Unified Gas Kinetic Scheme (DUGKS)
\cite{guo13}, the discrete Boltzmann models \cite{xu18}
and the lattice Boltzmann (LB) models \cite{ambrus16jcp}. 

Close to the hydrodynamic regime, 
reliable results can be obtained using lattice Boltzmann models based on 
the D3Q27 model \cite{yudistiawan10}, the spherical decomposition 
of the velocity space \cite{ambrus12pre} or the full-range 
Gauss-Hermite quadrature \cite{ambrus16jcp,meng13jfm}. 
As the degree of rarefaction increases, the number of velocities
required by such models to ensure accurate results increases significantly
\cite{ambrus16jcp,ambrus12pre,piaud14ijmpc,ambrus16jocs}. 
This is due to the discontinuity of the distribution function, which
develops due to the particle-wall interaction \cite{gross57,takata13}.
When the gas is far from equilibrium, this discontinuity can be managed
more efficiently with LB models based on half-range Gauss-Hermite 
quadratures. As demonstrated in the context of the Couette flow 
between parallel plates \cite{ambrus16jcp,ambrus16jocs}, the ratio between 
the number of velocities used when full-range or half-range Gauss-Hermite 
quadratures are employed on the Cartesian axis perpendicular to the wall 
increases dramatically for values of the Knudsen number ${\rm Kn}$
exceeding $0.1$ (e.g., at ${\rm Kn} = 0.5$, this ratio is $9.5$).

In this contribution, we validate our LB simulation results by comparison
to the DSMC results reported in Ref.~\cite{sharipov18}, which
were obtained for 
dilute gases comprised of ${}^3{\rm He}$ and ${}^4{\rm He}$ atoms
that interact via quantum scattering cross-sections, computed 
using ab initio potentials. The connection between the 
Shakhov model employed in the LB method and the interaction model 
employed in the DSMC method is made by implementing the relaxation 
time $\tau$ and the Prandtl number ${\rm Pr}$ such that the 
viscosity $\mu$ and heat conductivity $\kappa$ match the values 
computed in Ref.~\cite{cencek12}.

The paper is structured as follows. First, the application of the 
Shakhov model with reduced distributions for the simulation of gases 
with interparticle interactions based on ab initio potentials is discussed. 
Next, we introduce the mixed quadrature LB models, which employ 
the half-range Gauss-Hermite quadrature on the axis perpendicular to the wall.
The comparison of the numerical results obtained using the 
Shakhov collision model and the full DSMC analysis is further discussed. 
Finally, we present our conclusions.

\vspace{-2pt}

\section{SHAKHOV KINETIC MODEL FOR THE COUETTE FLOW}

\vspace{-2pt}

In this paper, we focus on the study of the Couette flow between parallel plates. 
The coordinate system is chosen such that the $x$ axis is perpendicular to the walls. The origin 
of the coordinate system is taken to be on the channel centerline, such that the 
left and right walls are located at $x = -L/2$ and $x = L/2$, respectively.
The plates are set in motion along the $y$ axis and the flow is studied in the 
Galilean frame where the left and right plates move with velocities $-u_w$ and $u_w$, respectively. 
Both plates are kept at constant temperature $T_w$.
In this case, the Boltzmann equation with the Shakhov approximation for the collision 
term can be written as follows \cite{shakhov68a,ambrus12pre,ambrus18pre}:
\begin{equation}
 \frac{\partial f}{\partial t} + \frac{p_x}{m} \partial_x f = -\frac{1}{\tau}\left[ 
 f - \feq(1 + \mathbb{S})\right],\label{eq:boltz}
\end{equation}
where $f$ is the particle distribution function, $p_x$ is the 
particle momentum 
along the direction perpendicular to the walls, $m$ is the 
particle mass and $\tau$ is the relaxation time.
The Maxwell-Boltzmann distribution function $\feq$
for the ideal gas is:
\vspace{-7pt}
\begin{equation}
 \feq = n g(p_x, u_x, T) g(p_y, u_y, T) g(p_z, u_z, T), \qquad 
 g(p, u, T) = \frac{1}{\sqrt{2\pi m K_B T}} e^{-(p - mu)^2 / 2mK_B T},
 \label{eq:feq}
\end{equation}
where $u_\alpha$ is the macroscopic velocity along the $\alpha$ direction 
($\alpha \in \{x, y, z\}$), while $n$ and $T$ are the particle number 
density and the local temperature, respectively.
The Shakhov term $\mathbb{S}$ is given by:
\begin{equation}
 \mathbb{S} = \frac{1-{\rm Pr}}{n K_B^2 T^2}
 \left(\frac{\bm{\xi}^{2}}{5mK_B T}-1\right)
 \bm{q} \cdot \bm{\xi},\label{eq:sdef} 
\vspace{-5pt}
\end{equation}
where $\xi_\alpha = p_\alpha -mu_\alpha$ and $q_\alpha$ are the $\alpha$ 
components of the peculiar velocity and heat flux vectors. 
The Prandtl number ${\rm Pr} = c_p \mu /\kappa$, where 
$c_p = 5 K_B/ 2m$, 
represents a free parameter of the Shakhov model. This
parameter can 
be used to tune the heat conductivity $\kappa$, while the 
viscosity is proportional to 
the relaxation time $\tau$ and is given by $\mu = \tau P$, 
where $P = n K_B T$ is the ideal gas pressure.

Since the flow properties are trivial along the $z$ direction,
it is convenient to introduce the
reduced distributions $\phi$ and $\chi$ through
\cite{li04,meng13jcp}:
% \vspace{-8pt}
\begin{equation}
 \phi = \int_{-\infty}^{\infty} dp_z \, f, \qquad 
 \chi = \int_{-\infty}^{\infty} dp_z \frac{p_z^2}{m}\,f.
\end{equation}
The equations obeyed by the reduced distributions $\phi$ and $\chi$ can be 
obtained by multiplying Eq.~\eqref{eq:boltz} with $1$ and $p_z^2/m$ and integrating 
over $p_z$:
% \vspace{-8pt}
\begin{equation}
 \frac{\partial}{\partial t} \left(
 \begin{array}{c}
  \phi \\ \chi
 \end{array}\right) + \frac{p_x}{m} \frac{\partial}{\partial x} \left(
 \begin{array}{c}
  \phi \\ \chi 
 \end{array}\right) = -\frac{1}{\tau}\left[
 \begin{array}{c}
  \phi - \phi^{\rm (eq)} (1 + \mathbb{S}_\phi) \\
  \chi - \chi^{\rm (eq)} (1 + \mathbb{S}_\chi) 
 \end{array}\right],\label{eq:boltz_red} 
\end{equation}
where $\phi^{\rm (eq)} = n g(p_x,u_x,T) g(p_y,u_y, T)$, $\chi^{\rm (eq)} = K_B T \phi^{\rm (eq)}$ 
and the Shakhov terms $\mathbb{S}_\phi$ and $\mathbb{S}_{\chi}$ are given by:
\begin{equation}
 \mathbb{S}_\phi = \frac{1-{\rm Pr}}{nK_B^2 T^2}
\left(\frac{\xi_x^{2} + \xi_y^2}{5mK_B T}-\frac{4}{5}\right) (q_x \xi_x + q_y \xi_y),\qquad
 \mathbb{S}_\chi = \frac{1-{\rm Pr}}{nK_B^2 T^2}
 \left(\frac{\xi_x^{2} + \xi_y^2}{5mK_B T}-\frac{2}{5}\right)  (q_x \xi_x + q_y \xi_y).
\end{equation}
The macroscopic quantities describing the fluid can be obtained as moments 
of $\phi$ and $\chi$, as follows:
% \vspace{-5pt}
\begin{eqnarray}
 \left(\begin{array}{c}
  n \\ u_i \\ T_{ij} 
 \end{array}\right) = \int dp_x dp_y \left(
 \begin{array}{c}
  1 \\ p_i / \rho \\ \xi_i \xi_j / m
 \end{array}\right) \phi,\qquad 
 T_{zz} = \int dp_x dp_y\, \chi, \\
 \left(
 \begin{array}{c}
  q_x \\ q_y 
 \end{array}\right) = \int dp_x dp_y \left[\left(
 \begin{array}{c}
 (\xi_x^2 + \xi_y^2) \xi_x /2m^2\\
 (\xi_x^2 + \xi_y^2) \xi_y /2m^2
 \end{array}\right) \phi + \left(
 \begin{array}{c}
  \xi_x/2m \\ \xi_y/2m
 \end{array}\right) \chi\right],
\end{eqnarray}
where $i, j$ take values in $\{x, y\}$, while 
$u_z = T_{xz} = T_{yz} = q_z = 0$. The temperature 
is obtained via $T = \frac{1}{3n K_B}(T_{xx} + T_{yy} + T_{zz})$.

Due to the symmetries of Eq.~\eqref{eq:boltz_red},
$\phi(-x, -p_x, -p_y) = \phi(x, p_x, p_y)$ and
$\chi(-x, -p_x, -p_y) = \chi(x, p_x, p_y)$ for $0 \le x \le L/2$,
such that only the right half of the channel can be considered.
At $x = L/2$, diffuse reflection boundary conditions are imposed:
\vspace{-6pt}
\begin{eqnarray}
 \phi(x = L/2, p_x < 0) &=& n_w g(p_x, 0, T_w) g(p_y, u_w, T_w),\nonumber\\
 \chi(x = L/2, p_x < 0) &=& n_w K_B T_w g(p_x, 0, T_w) g(p_y, u_w, T_w),
 \label{eq:diffuse}
\end{eqnarray}
where $n_{w}$ is obtained by imposing zero mass flux through the wall:
\begin{equation}
 n_{w} = -\frac{\displaystyle \int_{0}^\infty dp_x \int_{-\infty}^\infty dp_y \, 
 \phi(x = L/2)\, p_x}
 {\displaystyle \int_{-\infty}^0 dp_x \int_{-\infty}^\infty dp_y \, 
 g(p_x, 0,T_w) g(p_y, u_w, T_w) p_x}.
 \label{eq:diffuse_nw}
\end{equation}

\begin{figure}
% \begin{center}
\begin{tabular}{ccc}
\includegraphics[angle=0,width=0.32\linewidth]{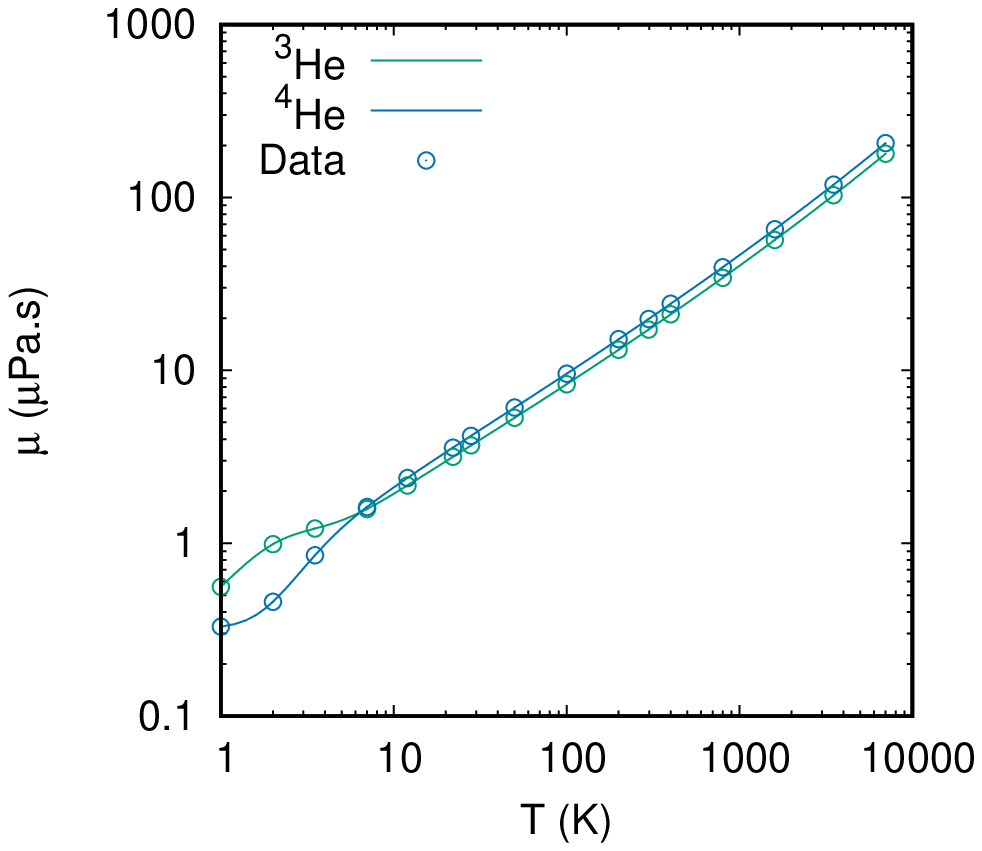} &
\includegraphics[angle=0,width=0.32\linewidth]{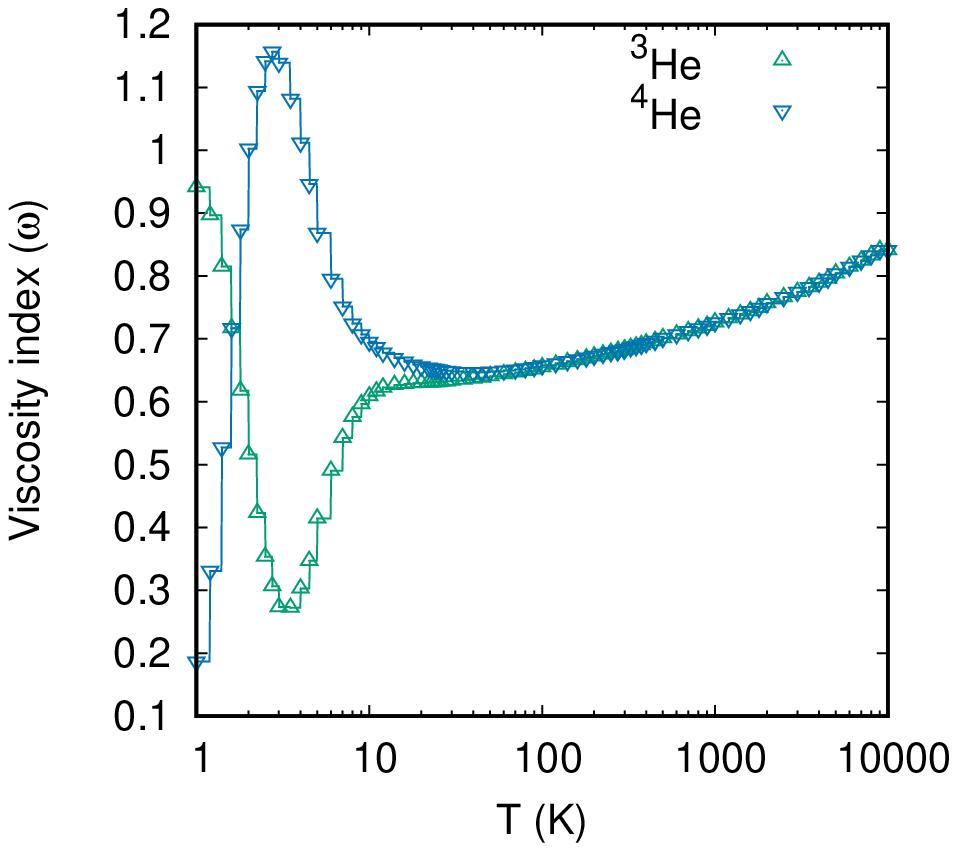} &
\includegraphics[angle=0,width=0.32\linewidth]{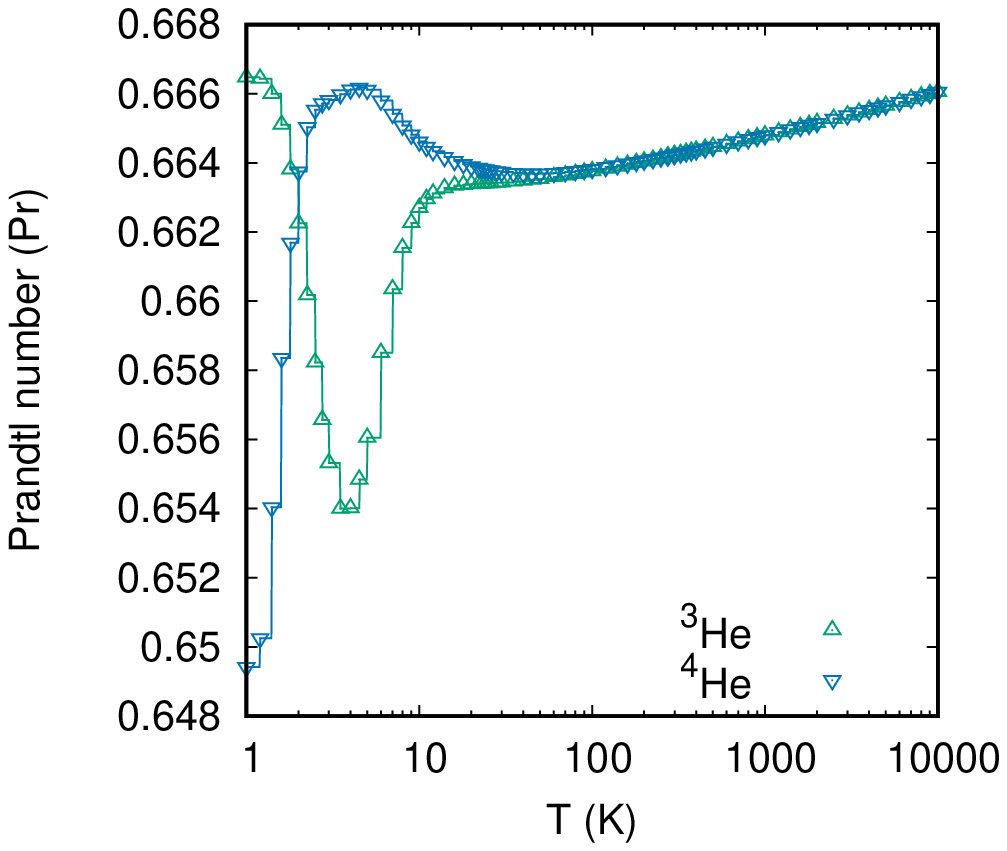} 
% \vspace{-10pt}
\end{tabular}
% \end{center} 
\caption{
Temperature dependence of (a) dynamic viscosity $\mu$ (in $\mu{\rm Pa} \cdot s$);
(b) viscosity index $\omega$; and (c) Prandtl number ${\rm Pr}$, as given by 
Eqs.~\eqref{eq:mun} and \eqref{eq:interp}. 
In (a), the tabulated values \cite{cencek12} for the viscosity 
are shown using hollow circles, while the continuous lines correspond to 
the piecewise functions defined in Eq.~\eqref{eq:mun}. 
The viscosity index and Prandtl number are shown as piecewise constant functions,
while the symbols mark the values obtained via 
the tabulated data from Ref.~\cite{cencek12}.
% \vspace{-12pt}
\label{fig:interp}}
\end{figure}

The connection between the relaxation time approximation of the Boltzmann 
equation and the full collision integral for a given interaction model can 
be made at the level of the transport coefficients. For simple interaction 
models such as the hard-sphere or Maxwell molecules gases, the viscosity 
has a temperature dependence of the form
% \vspace{-3pt}
\begin{equation}
 \mu = \mu_{\rm ref} (T / T_{\rm ref})^{\omega},
%  \vspace{-3pt}
 \label{eq:mu}
\end{equation}
where the viscosity index $\omega$ takes the values $1/2$ and $1$ for 
hard sphere and Maxwell molecules, respectively. For real gases, $\omega$ 
is in general temperature-dependent. In the contex of interactions based 
on ab initio potentials, the viscosity $\mu$ was tabulated for gases comprised 
of ${}^3{\rm He}$ and ${}^4{\rm He}$ atoms in the supplementary materials of 
Ref.~\cite{cencek12}.
In this work, we employ Eq.~\eqref{eq:mu} in a piecewise 
fashion by determining 
appropriate values of $\omega$ in order to interpolate the tabulated data, as follows.
Considering that the data table contains $N$ entries, let $\mu_n$ and $\mu_{n+1}$ 
($1 \le n < N$) represent two consecutive values of the viscosity, corresponding
to the values $T_n$ and $T_{n+1}$ of the temperature. For the interval 
$T \in [T_n, T_{n+1}]$, Eq.~\eqref{eq:mu} is replaced by:
\vspace{-7pt}
\begin{equation}
 \mu^{(n)}(T) = \mu_n (T / T_n)^{\omega_n}, \qquad
 \omega_n = \frac{\ln(\mu_{n+1} / \mu_n)}{\ln(T_{n+1} /T_n)},
 \vspace{-5pt}
 \label{eq:mun}
\end{equation}
such that $\mu^{(n)}(T_n) = \mu_n$ and $\mu^{(n)}(T_{n+1}) = \mu_{n+1}$.
We also take advantage of the freedom in controlling the Prandtl 
number ${\rm Pr}$ via the Shakhov collision model in order to track the 
(small) variations of ${\rm Pr}$ with temperature. 
For simplicity, we consider a piecewise constant implementation of ${\rm Pr}$,
such that ${\rm Pr} = {\rm Pr}_n$ when $T_n \le T < T_{n+1}$, where
${\rm Pr}_n = c_p \mu_n /\kappa_n$ is obtained using the 
tabulated value $\kappa_n$ of the heat conductivity corresponding to 
$T = T_n$.
In order to access temperature ranges outside the tabulated data, 
Eq.~\eqref{eq:mu} is extrapolated by taking $\mu(T) = \mu^{(1)}(T)$ 
and ${\rm Pr}(T) = {\rm Pr}_1$ for $T < T_2$, while $\mu(T) = \mu^{(N - 1)}(T)$ 
and ${\rm Pr}(T) = {\rm Pr}_N$ for $T > T_{N}$. The 
mathematical expression for the above algorithm is:
\begin{equation}
 \mu(T) = \left\{
 \begin{array}{lll}
  \mu^{(1)}(T), & \quad& T < T_2, \\
  \mu^{(n)}(T), & \quad& T_n < T < T_{n+1}, \\
  \mu^{(N-1)}(T), & \quad& T_{N} < T,
 \end{array}\right. \qquad 
 {\rm Pr}(T) = \left\{
 \begin{array}{lll}
  {\rm Pr}_1, & \quad& T < T_2, \\
  {\rm Pr}_n, & \quad& T_n < T < T_{n+1}, \\
  {\rm Pr}_{N}, & \quad& T_{N} < T,
 \end{array}\right.
 \label{eq:interp}
\end{equation}
where $n = 2, 3, \dots N -1$ refers to the index of the tabulated values in 
Ref.~\cite{cencek12}.
The interpolation corresponding to Eq.~\eqref{eq:interp} is shown in 
Fig.~\ref{fig:interp}, where the viscosity $\mu(T)$ is shown as a continuous 
function, while the viscosity index $\omega$ and the Prandtl number 
${\rm Pr}$ are shown as piecewise functions. It can be seen that while ${\rm Pr}$ 
is confined within a few percent of the expected value $2/3$, the 
viscosity index presents significant variations, expecially in the low 
temperature regime.

The degree of rarefaction of the flow can be described using the rarefaction 
parameter $\delta$, defined through \cite{sharipov18}:
\vspace{-3pt}
\begin{equation}
 \delta = \frac{L P_{\rm ref}}{\mu(T_w) v_{\rm ref} \sqrt{2}},
 \label{eq:delta}
 \vspace{-5pt}
\end{equation}
where $P_{\rm ref} = n_{\rm ref} K_B T_w$, $n_{\rm ref}$ is the average 
particle number densiy and $v_{\rm ref} = \sqrt{K_B T_w / m}$ is the 
reference speed. 
The relaxation time can thus be written in terms of the 
rarefaction parameter as follows:
\vspace{-5pt}
\begin{equation}
 \tau = \frac{\mu(T) / \mu(T_w)}{(n/n_{\rm ref}) (T/T_w)} 
 \frac{t_{\rm ref}}{\delta \sqrt{2}},\label{eq:tau}
 \vspace{-10pt}
\end{equation}
where the reference time is $t_{\rm ref} = L / v_{\rm ref}$.

\vspace{-2pt}

\section{MIXED QUADRATURE LATTICE BOLTZMANN MODELS}

\vspace{-2pt}

In this section, the LB algorithm employed to solve Eq.~\eqref{eq:boltz_red}
is briefly described. Our implementation is based on the concept of mixed 
quadratures 
\cite{ambrus16jcp,ambrus16jocs,gibelli12}, which allows the quadrature to
 be controlled on 
each axis independently (more details on the concept of Gaussian quadrature 
can be found in, e.g., Refs.~\cite{hildebrand87,shizgal15}).
In particular, the half-range Gauss-Hermite quadrature is employed on the 
$x$ axis, where the distribution function becomes discontinuous due to the 
diffuse reflection boundary conditions \cite{gross57,takata13}. On the
periodic ($y$) direction, the full-range Gauss-Hermite quadrature is
employed since there are 
no discontinuities of the distribution function with respect to $p_y$. 
The technical details regarding the construction of such models are given 
in Ref.~\cite{ambrus16jcp} in the context of the 2D Couette flow and 
the application of these models to the 3D Couette flow using reduced 
distributions is discussed in Ref.~\cite{ambrus18pre}.
In this section, the main ingredients necessary to employ these models 
are summarized.

The half-range Gauss-Hermite quadrature allows the recovery of the 
integrals with respect to $p_x$ on each semi-axis individually, as follows:
\vspace{-5pt}
\begin{equation}
 \int_0^\infty \frac{dp_x}{\sqrt{2\pi}} e^{-p_x^2/2p_{0,x}^2} p_x^s \simeq 
 p_{0,x} \sum_{i = 1}^{Q_x} w_i^{\hh} p_{x,i}^s, \qquad 
 \int_{-\infty}^0 \frac{dp_x}{\sqrt{2\pi}} e^{-p_x^2/2p_{0,x}^2} p_x^s \simeq
 p_{0,x} \sum_{i = 1}^{Q_x} w_i^{\hh} (-p_{x,i})^s,
 \vspace{-5pt}
\end{equation}
where equality holds when $2Q_x > s$. The ratios between the discrete 
momentum values $p_{x,i}$ ($1 \le i \le Q_x$) 
and the arbitrary reference momentum scale 
$p_{0,x}$ are the roots of the half-range Hermite polynomial
$\hh_{Q_x}(z)$, of 
order $Q_x$ [i.e. $\hh_{Q_x}(p_{x,i}/p_{0,x}) = 0$]. The quadrature 
weight corresponding to the quadrature point $p_{x,i} / p_{0,x}$ is
obtained through \cite{ambrus16jcp}:
\begin{equation}
 w_i^{\hh} = \frac{p_{x,i} a_{Q_x-1}^2}{\hh^2_{Q_x -1}(p_{x,i}/p_{x,0}) 
[p_{x,i} + p_{0,x} \hh_{Q_x}^2(0) / \sqrt{2\pi}]},
\end{equation}
where $a_\ell = \hh_{\ell+1,\ell+1} / \hh_{\ell,\ell}$ and
$\hh_{\ell,s}$ is the coefficient of $z^s$ in $\hh_\ell(z)$.
We use the convention that $p_{x,i+Q_x} = -p_{x,i}$, such that
$w^{\hh}_{x,i+Q_x} = w^{\hh}_{x,i}$.

The integrals with respect to $p_y$ can be recovered using 
the full-range Gauss-Hermite quadrature, as follows:
\begin{equation}
 \int_{-\infty}^\infty \frac{dp_y}{\sqrt{2\pi}} e^{-p_y^2/2 p_{0,y}^2} p_y^s \simeq
 p_{0,y} \sum_{j = 1}^{Q_y} w_j^H p_{y,j}^s, \label{eq:boltzk}
\end{equation}
where equality is achieved when the quadrature order 
$Q_y$ satisfies $2Q_y > s$. The ratios 
$p_{y,j} / p_{0,y}$ are the roots of the full-range Hermite 
polynomial $H_{Q_y}(z)$ of order $Q_y$, i.e. $H_{Q_y}(p_{y,j}/p_{0,y}) = 0$,
where $p_{0,y}$ is an arbitrary reference momentum scale.
The quadrature weights $w^H_j$ can be computed using \cite{ambrus16jcp}:
\begin{equation}
 w_j^H = \frac{Q_y!}{H^2_{Q_y+1}(p_{y,j}/p_{0,y})}.
\end{equation}

After the discretization of the momentum space, 
the factors $g(p_x,u_x, T)$ and 
$g(p_y,u_y,T)$ in Eq.~\eqref{eq:feq} are replaced 
by a set of polynomial truncations $g_{x,i}(u_x, T)$ and $g_{y,j}(u_y, T)$
of orders $0 \le N_x < Q_x$ and $0 \le N_y < Q_y$ with respect to 
the half-range and the full-range Hermite polynomials, respectively. 
In particular, $g(p_x, u_x, T)$ is replaced by
\begin{equation}
 g_{x,i} = \frac{w_i^{\hh} \sqrt{2\pi}}{e^{-p_{x,i}^2/2p_{0,x}^2}} g(p_{x,i}, u_x, T) 
 = \frac{w_i^{\hh}}{2} \sum_{s = 0}^{N_x} 
 \left(\frac{m K_B T}{2p_{0,x}^2}\right)^{\frac{s}{2}} \left[(1 + {\rm erf}\, \zeta_{x,i})
 P_s^+(\zeta_{x,i})
 + \frac{2e^{-\zeta_{x,i}^2}}{\sqrt{\pi}} P_s^*(\zeta_{x,i})\right]
 \Phi_{s,i}^{N_x},
 \label{eq:gx}
\end{equation}
where $1 \le i \le 2Q_x$, $\zeta_{x,i} = \sigma_{x,i} u_x \sqrt{m/2K_B T}$,
$\sigma_{x,i}$ is the sign of $p_{x,i}$,
$\Phi_{s,i}^{N_x} = \sum_{\ell = s}^{N_x} \hh_{\ell,s} \hh_{\ell}(|p_{x,i}/p_{0,x}|)$
and the polynomials $P_s^+(\zeta)$ and $P_s^*(\zeta)$ are given by:
\vspace{-10pt}
\begin{equation}
 P_s^\pm(\zeta) = e^{\mp \zeta^2} \frac{d^s}{d\zeta^s} e^{\pm \zeta^2}, \qquad 
 P_s^*(\zeta) = \sum_{j = 0}^{s-1} 
 \left(\begin{array}{c}
  s \\ j        
 \end{array}\right) P_j^+(\zeta) P_{s-j-1}^-(\zeta).
 \vspace{-10pt}
\end{equation}
Similarly, $g(p_y,u_y,T)$ is replaced by:
\begin{equation}
 g_{y,j} = \frac{w_j^H \sqrt{2\pi}}{e^{-p_{y,j}^2/2p_{0,y}^2}} g_y(p_{y,j}, u_y, T)
 = w^H_j \sum_{\ell = 0}^{N_y} H_\ell(p_{y,j}) 
 \sum_{s = 0}^{\lfloor \ell/2\rfloor} 
 \frac{1}{2^s s! (\ell - 2s)!}
 \left(\frac{m K_B T}{p_{0,y}^2} - 1\right)^s 
 \left(\frac{mu_y}{p_{0,y}}\right)^{\ell-2s},
 \label{eq:gy}
\end{equation}
where $1 \le j \le Q_y$.
The polynomial truncations \eqref{eq:gx} and \eqref{eq:gy} are constructed such that 
the half-space and full-space moments of $g(p_x,u_x,T)$ and $g(p_y,u_y, T)$, respectively,
are exactly recovered via the following quadrature sums
\cite{ambrus16jcp,ambrus16jocs}:
\begin{eqnarray}
 \int_0^\infty dp_x g(p_x,u_x,T) p_x^s = \sum_{i = 1}^{Q_x} g_{x,i} p_{x,i}^s, \qquad
 \int_{-\infty}^0 dp_x g(p_x,u_x,T) p_x^s = \sum_{i = Q_x+1}^{2Q_x} g_{x,i} p_{x,i}^s,\nonumber\\
 \int_{-\infty}^\infty dp_y g(p_y,u_y,T) p_y^\ell = \sum_{j = 1}^{Q_y} g_{y,j} p_{y,j}^\ell,
 \hspace{90pt}
\end{eqnarray}
for all values of $s$ and $\ell$ which satisfy $0 \le s \le N_x$ and $0 \le \ell \le 
N_y$.

The resulting mixed quadrature LB models are denoted through
${\rm HHLB}(N_x; Q_x) \times {\rm HLB}(N_y; Q_y)$, 
where $N_x$ and $N_y$ are the expansion orders of the 
equilibrium distribution and $Q_x$ and $Q_y$ are the quadrature 
orders of the half-range and full-range Gauss-Hermite quadratures.

The solution of 
Eq.~\eqref{eq:boltz_red} is obtained using the total variation diminshing 
(TVD) third order Runge-Kutta (RK-3) integration method introduced in 
Ref.~\cite{shu88} together with the fifth order weighted essentially 
non-oscillatory (WENO-5) advection scheme introduced in Ref.~\cite{jiang96}.
In order to accurately capture the Knudsen layer in the vicinity of the 
wall at $x = L/2$, a coordinate stretching procedure is employed 
following Ref.~\cite{mei98jcph}, which is given through the following 
coordinate transformation \cite{ambrus18pre,busuioc19}:
 \vspace{-10pt}
\begin{equation}
 x(\eta) = \frac{L}{2A} \tanh\eta.\label{eq:stretch}
\end{equation}
The stretching parameter $A$ takes values between $0$ and $1$, while the 
domain for $\eta$ is $0 \le \eta \le {\rm arctanh}\, A$. The discretization
of the flow domain is performed using $S$ equidistant values for $\eta$, 
namely $\eta_s = \frac{1}{S} (s - 1/2) {\rm arctanh}\, A$ ($1 \le s \le S$).
This allows the advection scheme to be implemented using a finite
difference formulation, as described in Refs.~\cite{ambrus18pre,busuioc19}.
For simplicity, we only consider the case when $A = 0.98$ in the simulations 
presented in the following sections.

\vspace{-5pt}

\section{NUMERICAL RESULTS}

\vspace{-2pt}

In order to compare the DSMC and LB methods considered in this paper, 
we performed simulations for gases comprised of ${}^3{\rm He}$ and 
${}^4{\rm He}$ atoms, for wall temperatures varying 
between $1\ {\rm K}$ and $3000\ {\rm K}$. The numerical scheme of the 
DSMC method used in the present work is described in Ref.~\cite{sharipov18}.
The wall velocity is set to $u_w = \sqrt{2 K_B T_w / m}$.
Three values of the rarefaction parameter $\delta$ were used, namely
$\delta \in \{10, 1, 0.1\}$.  The number of nodes was always kept at 
$S = 16$ and the quadrature orders employed are discussed at the 
end of this section. 

\begin{figure}
\begin{tabular}{ccc}
\includegraphics[angle=0,width=0.31\linewidth]{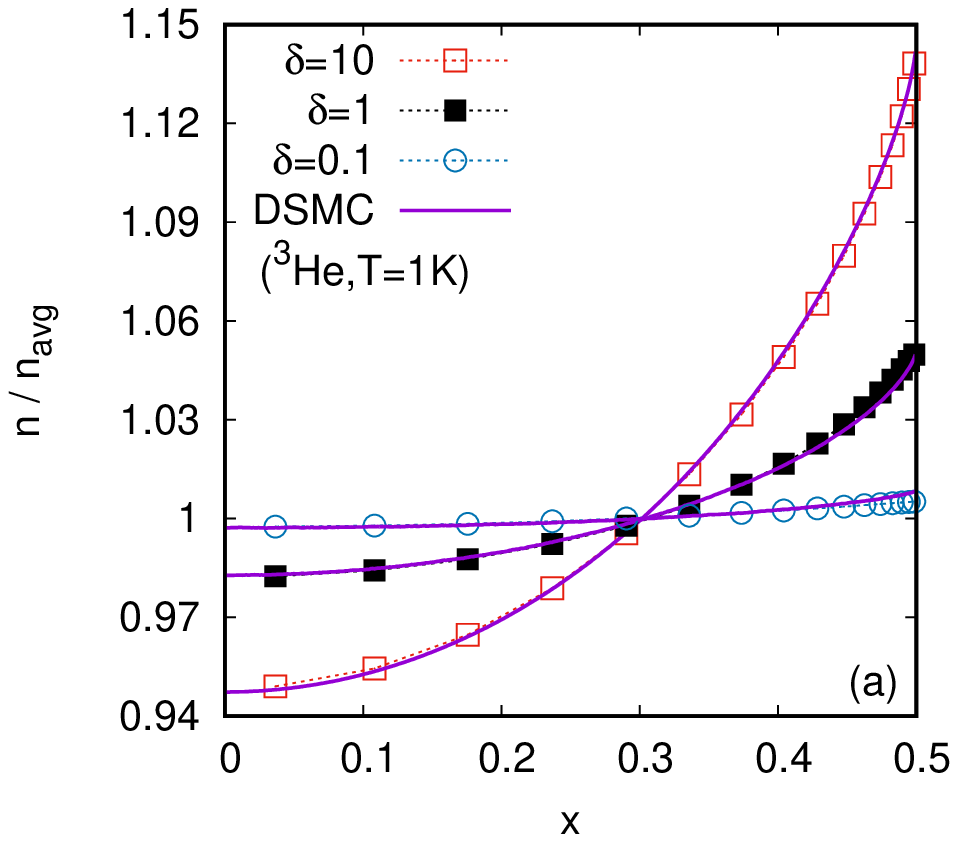} &
\includegraphics[angle=0,width=0.31\linewidth]{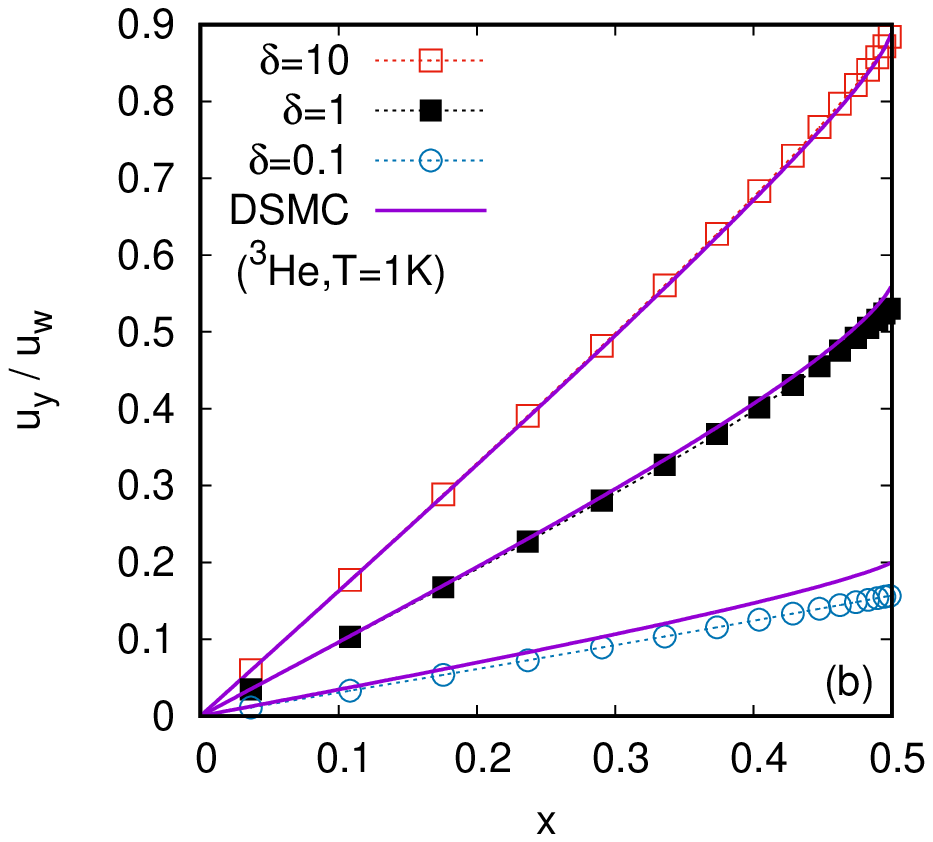} &
\includegraphics[angle=0,width=0.31\linewidth]{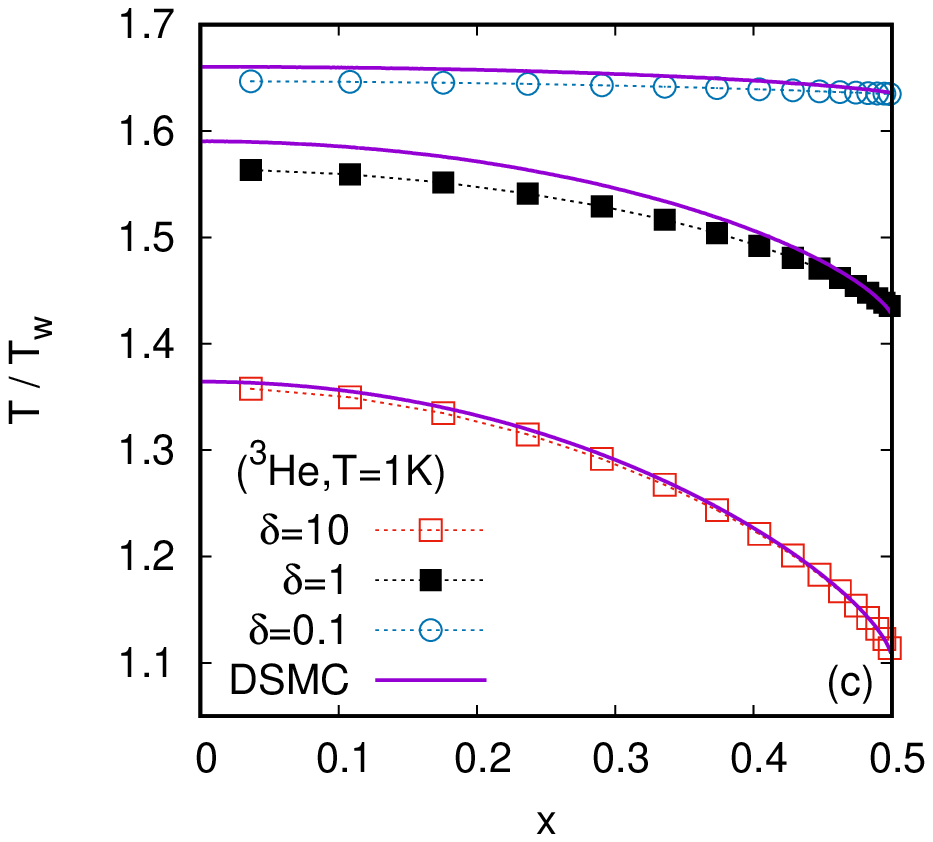} \\
\includegraphics[angle=0,width=0.31\linewidth]{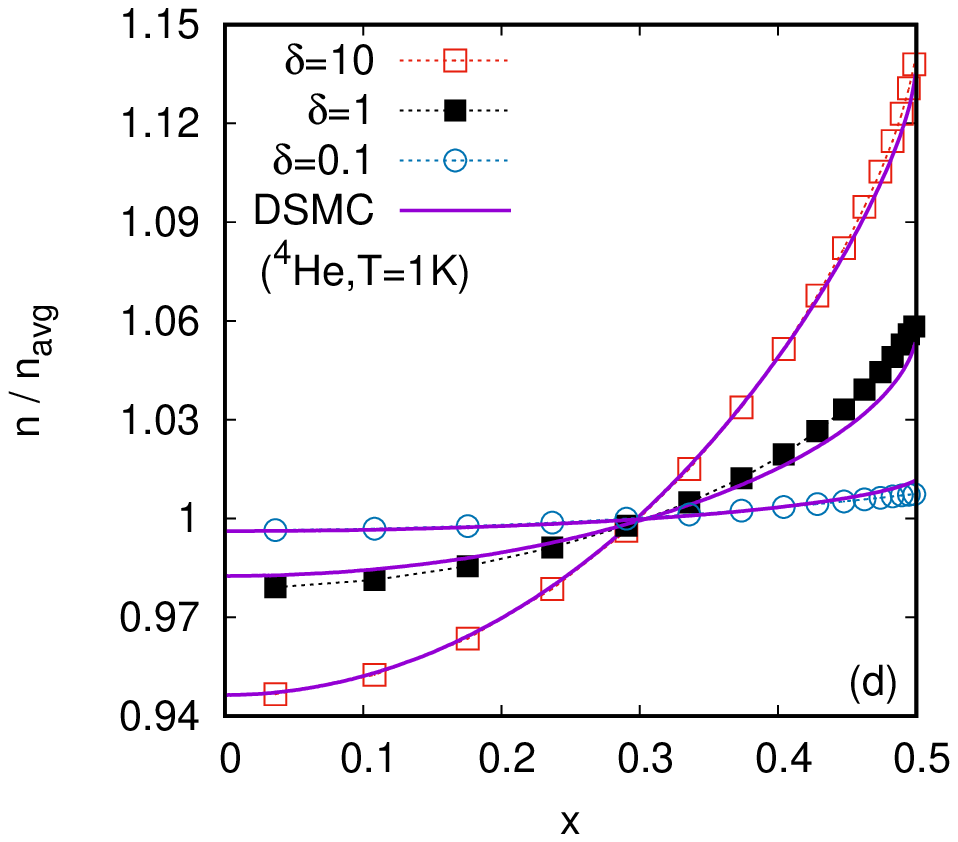} &
\includegraphics[angle=0,width=0.31\linewidth]{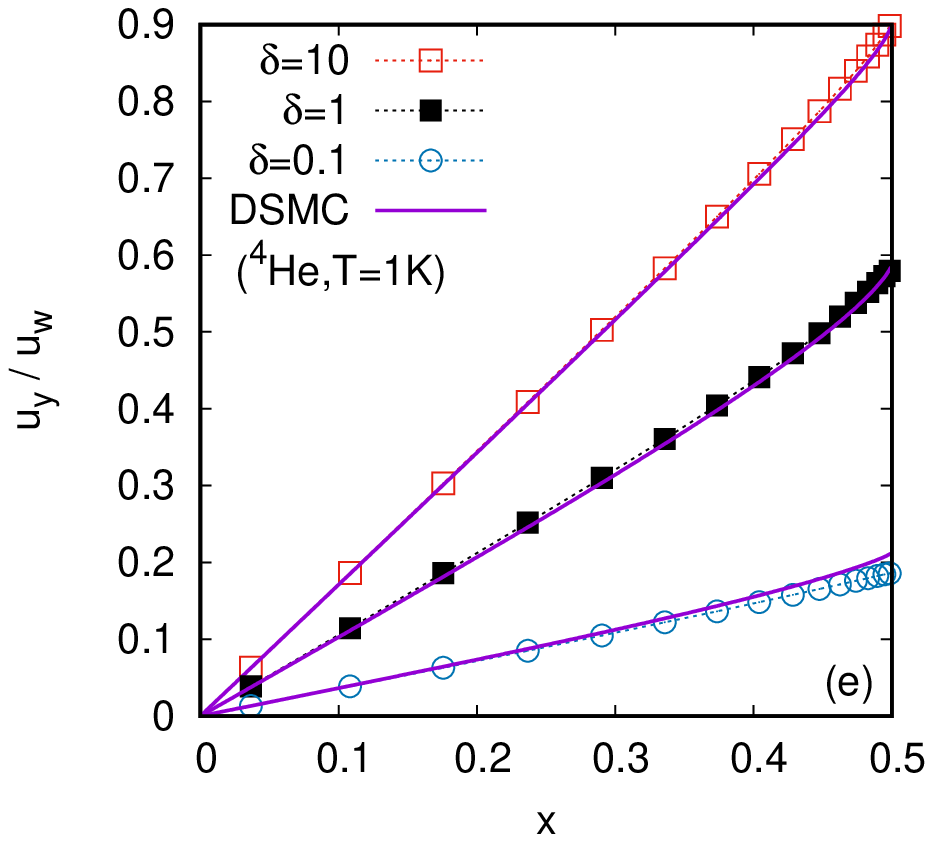} &
\includegraphics[angle=0,width=0.31\linewidth]{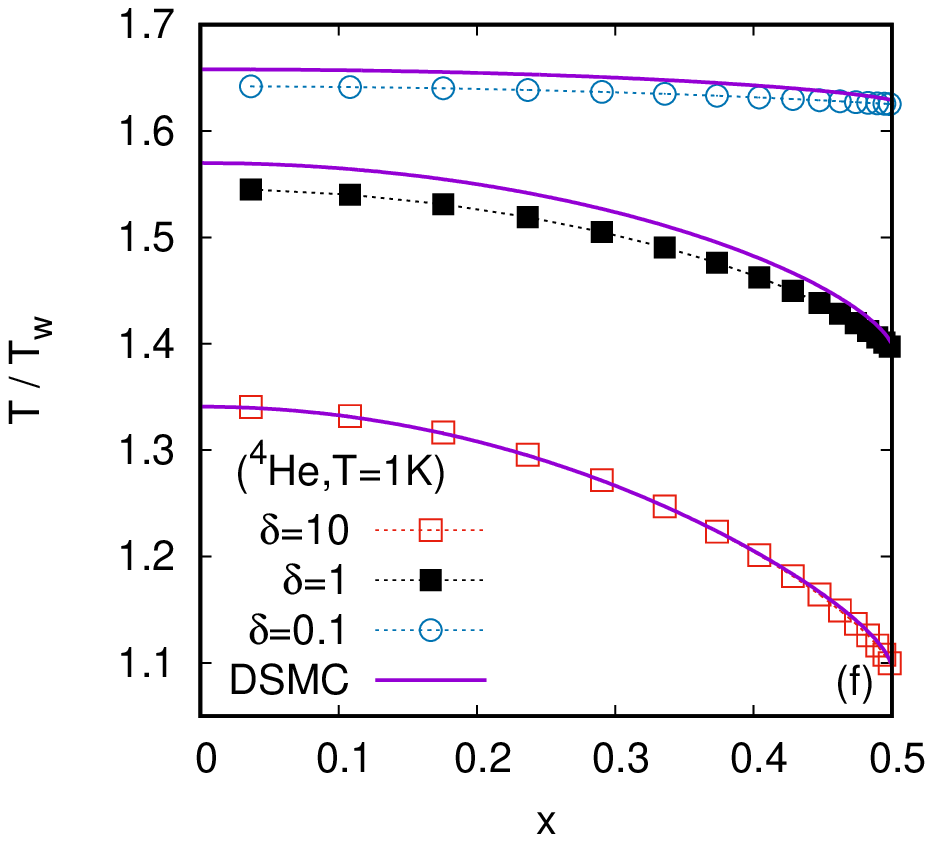} 
\end{tabular}
\caption{Comparison between the LB (dotted lines and points) and 
DSMC (continuous lines) results for the profiles of $n$ (left), 
$u_y$ (middle) and $T$ (right) through half of the channel 
($0 \le x\le L/2$), for ${}^3{\rm He}$ (top) and 
${}^4{\rm He}$ (bottom) gas constituents.
The wall temperature is set to $T_w = 1\ {\rm K}$ and 
the wall velocity is $u_w = \sqrt{2 K_B T_w / m}$.
\label{fig:T1}}
\end{figure}

\begin{figure}
\begin{tabular}{ccc}
\includegraphics[angle=0,width=0.31\linewidth]{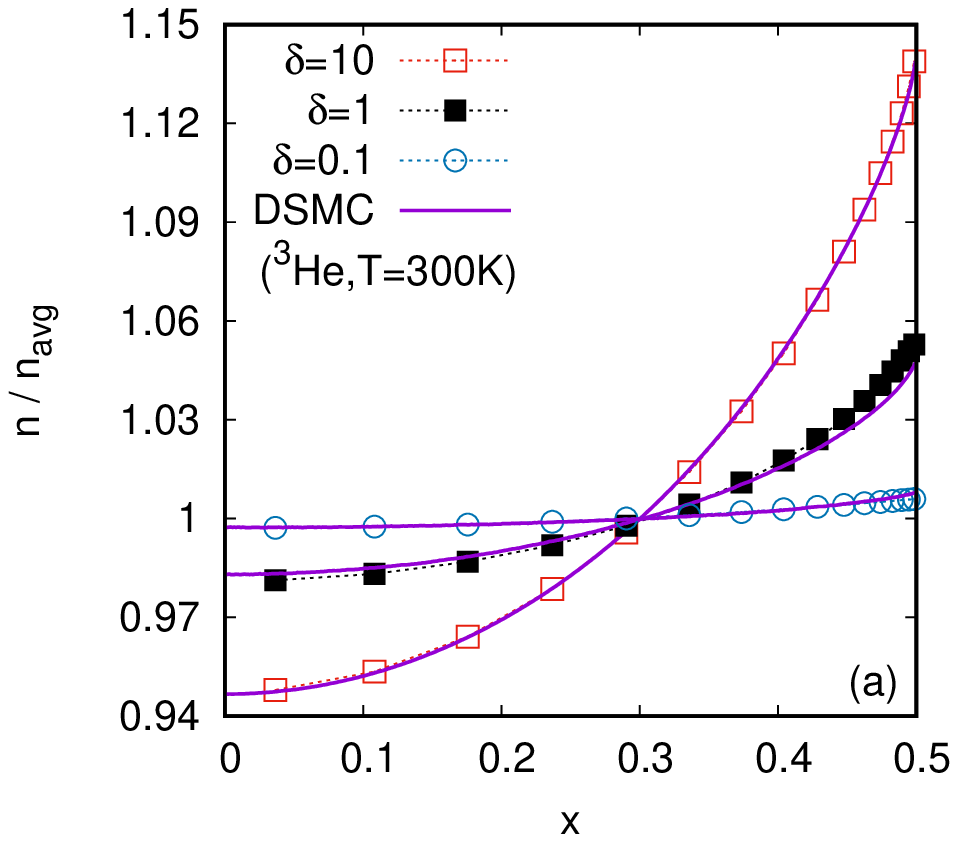} &
\includegraphics[angle=0,width=0.31\linewidth]{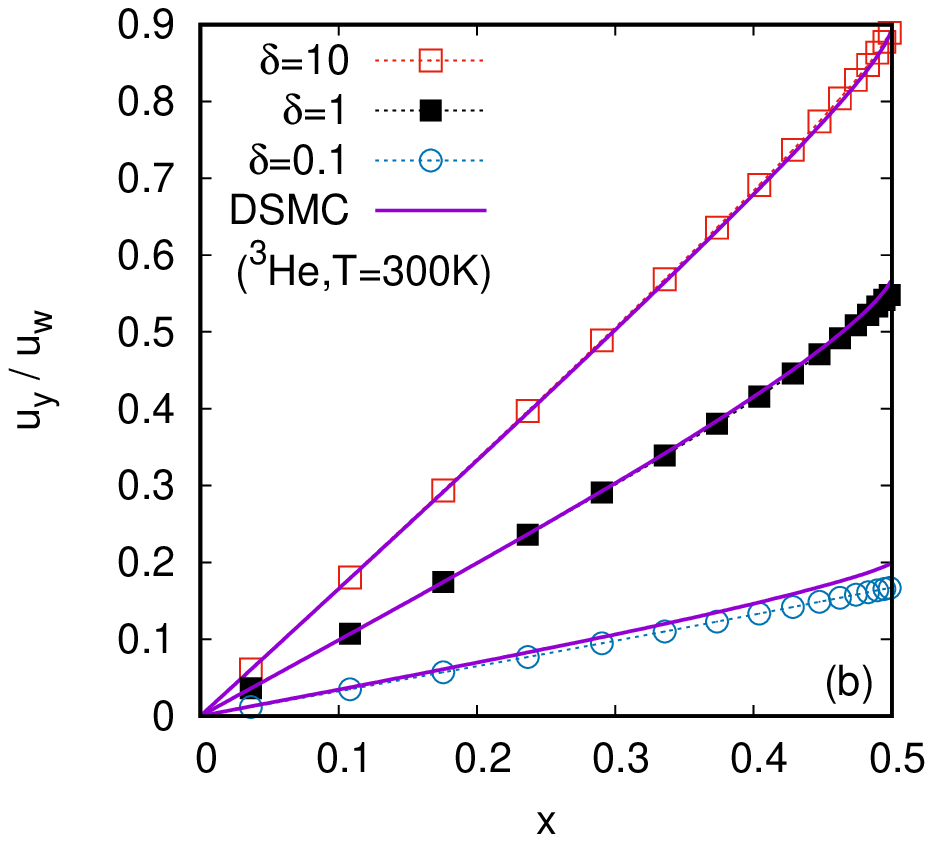} &
\includegraphics[angle=0,width=0.31\linewidth]{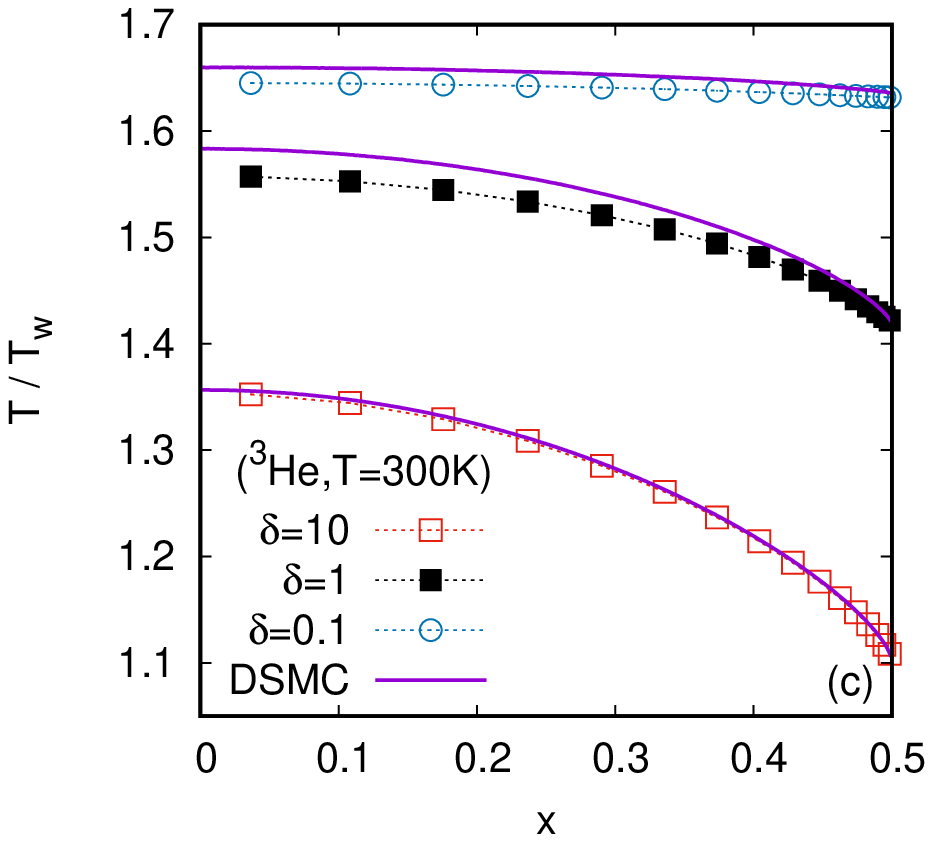} \\
\includegraphics[angle=0,width=0.31\linewidth]{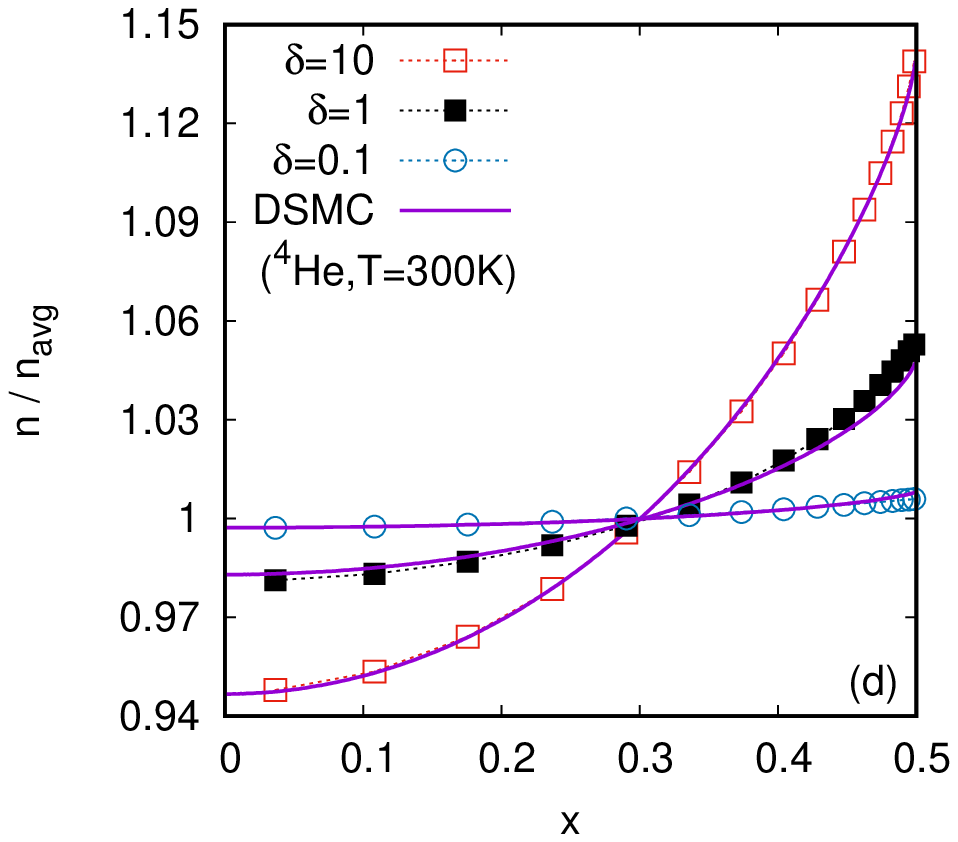} &
\includegraphics[angle=0,width=0.31\linewidth]{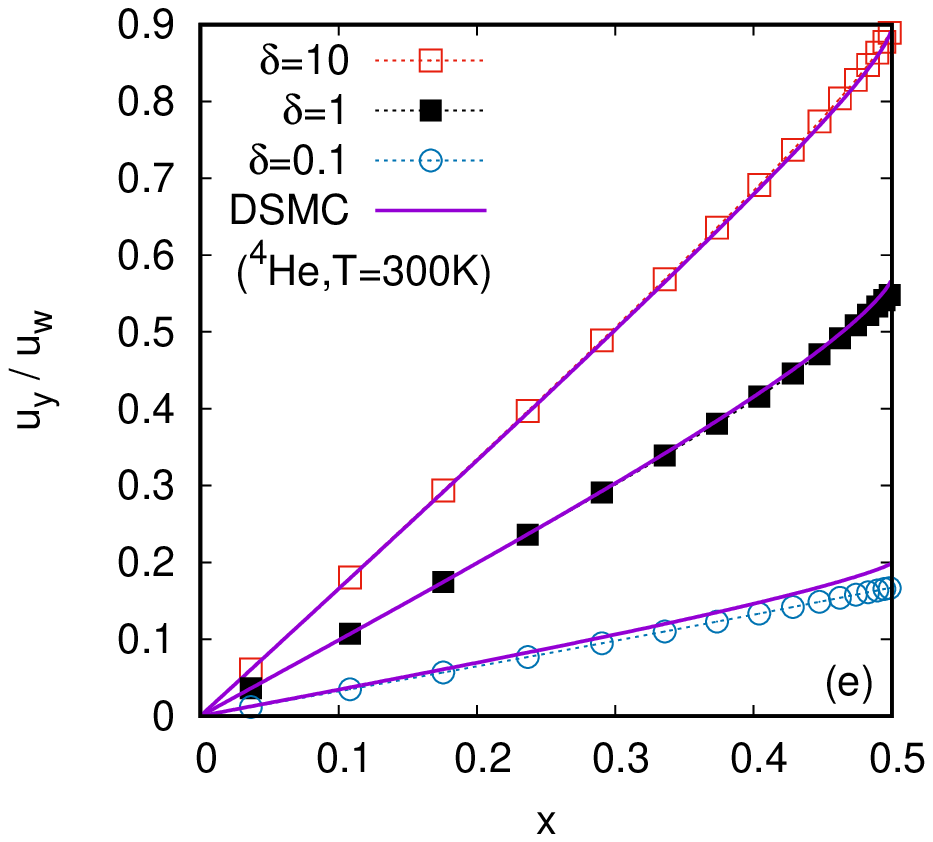} &
\includegraphics[angle=0,width=0.31\linewidth]{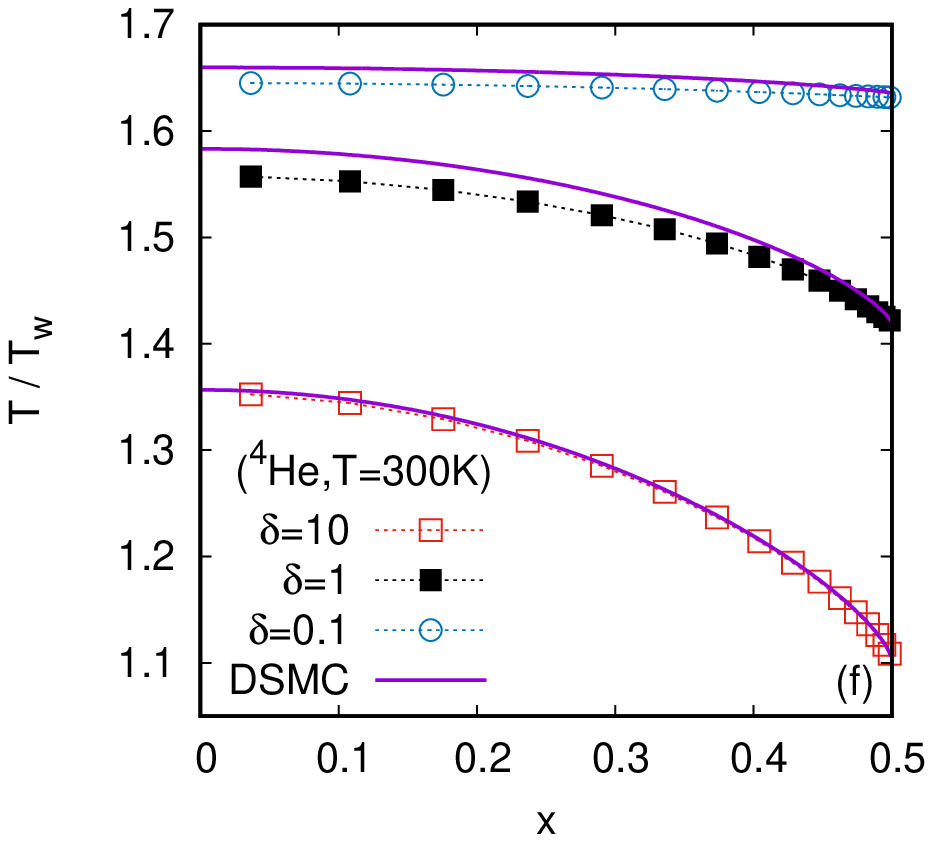} 
\end{tabular}
\caption{The same as in Fig.~\ref{fig:T1}, but with wall temperature 
$T_w = 300\ {\rm K}$.\label{fig:T300}}
\end{figure}

Figures~\ref{fig:T1} and \ref{fig:T300} show a comparison of the LB and DSMC 
results for $T_w = 1\ {\rm K}$ and $300\ {\rm K}$ at the level 
of the profiles of $n$, $u_y$ and $T$, for both the
${}^3{\rm He}$ and the ${}^4{\rm He}$ gases. Good agreement can be seen at 
$\delta = 10$, while at $\delta = 0.1$, there are some discrepancies
between the results for $u_y$ and $T$. The discrepancy in the temperature 
profile persists also at $\delta = 1$. 

\begin{figure}
\begin{tabular}{cc}
\includegraphics[angle=0,width=0.45\linewidth]{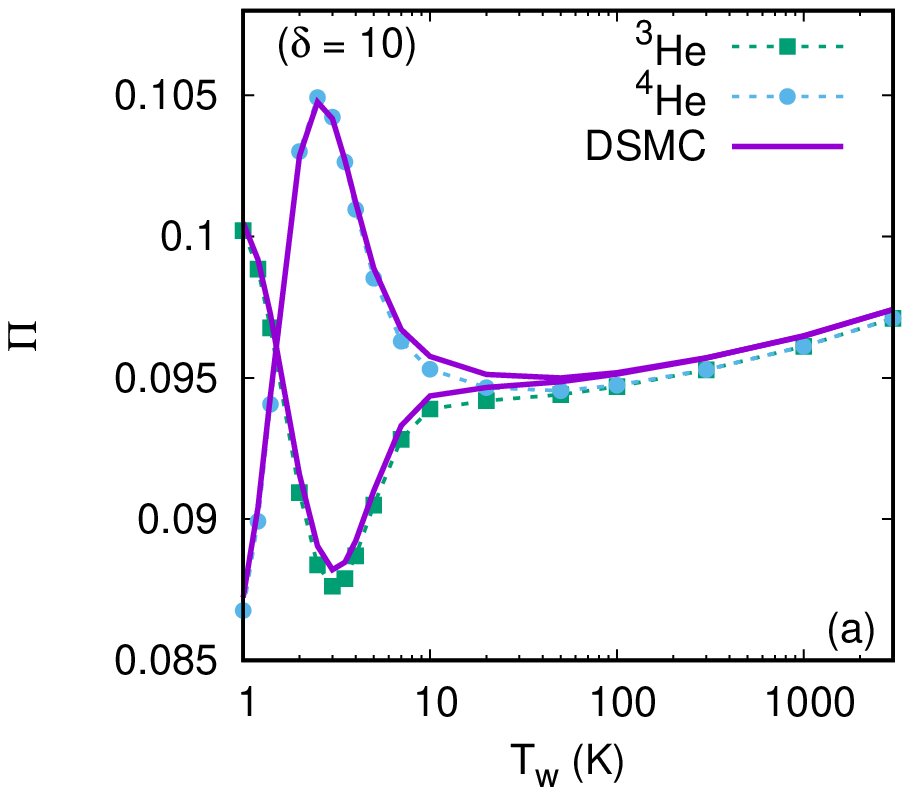} &
\includegraphics[angle=0,width=0.45\linewidth]{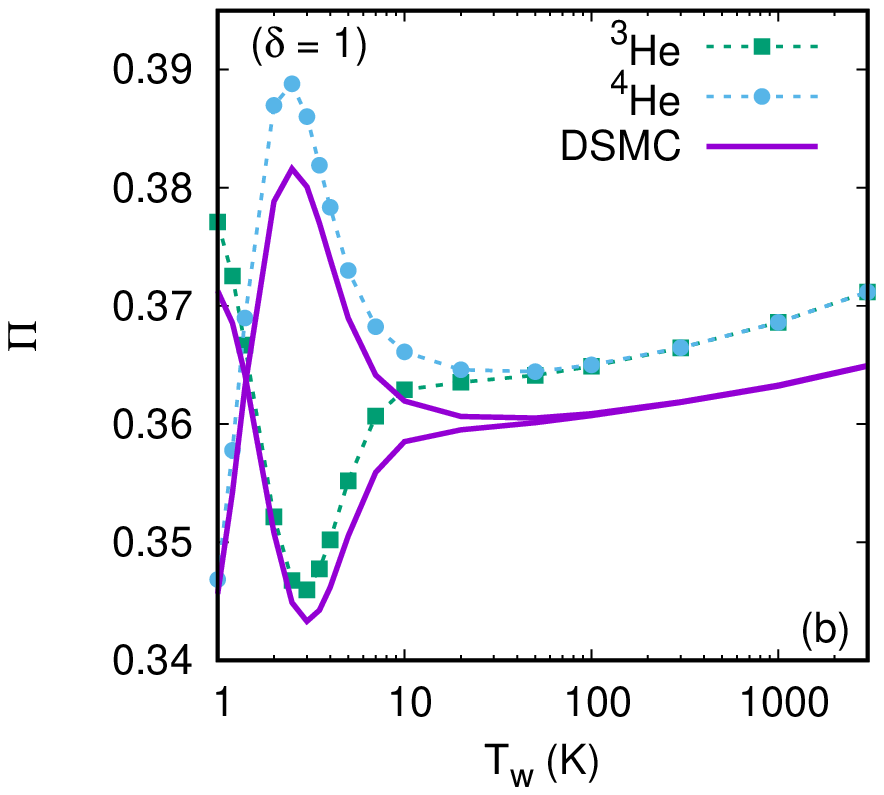} \\
\includegraphics[angle=0,width=0.45\linewidth]{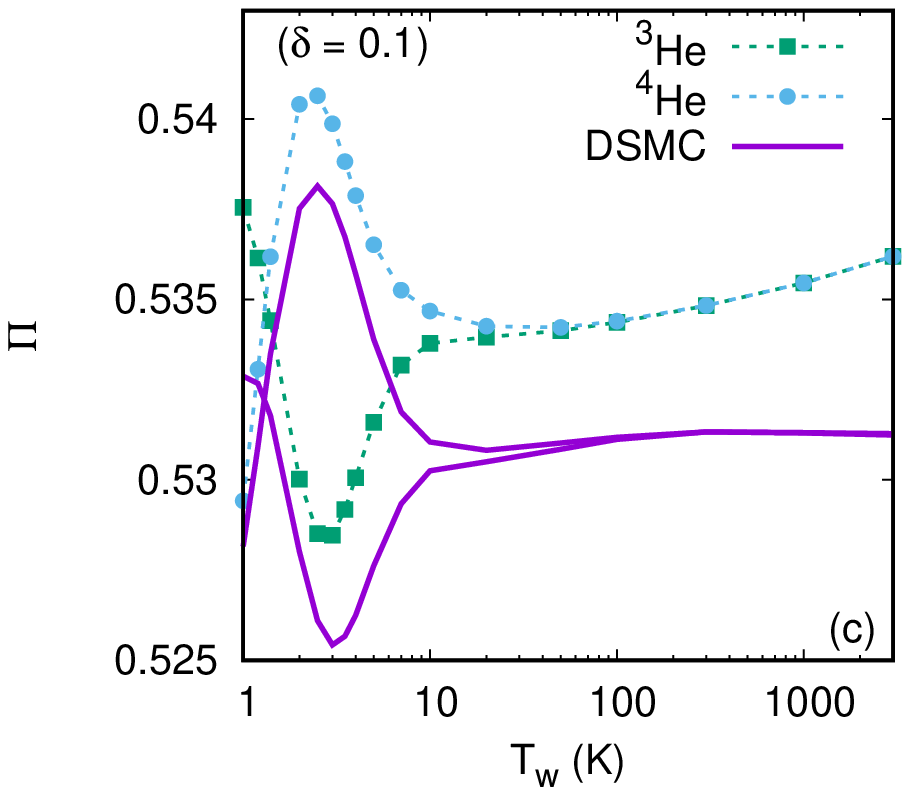} &
\includegraphics[angle=0,width=0.45\linewidth]{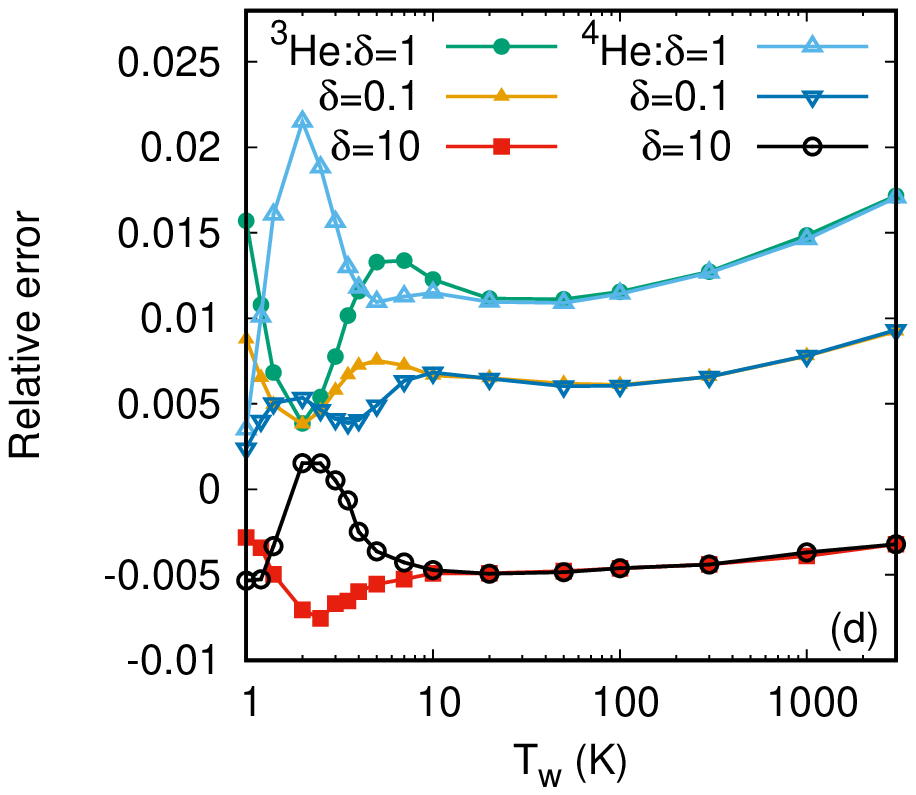} 
\end{tabular}
\caption{(a-c) Dependence of $\Pi$ \eqref{eq:Pi} on the wall temperature $T_w$ 
for both ${}^3{\rm He}$ and ${}^4{\rm He}$, at $\delta = 10$, $1$ and $0.1$. 
(d) Relative error $\Pi_{\rm LB} / \Pi_{\rm DSMC} - 1$ as a function of 
$T_w$ for $\delta \in \{10, 1, 0.1\}$.\label{fig:Pi}}
\end{figure}

We next consider a comparison of the LB and DSMC results for the 
off-diagonal component of the stress tensor $T_{xy}$, which we present 
through the non-dimensional number 
\begin{equation}
 \Pi = -\frac{T_{xy} v_{\rm ref}}{P_{\rm ref} u_w \sqrt{2}}.\label{eq:Pi}
\end{equation}
Noting that $u_w = \sqrt{2K_B T_w / m} = v_{\rm ref} \sqrt{2}$,
it can be seen that $\Pi = -T_{xy} / 2P_{\rm ref}$.
Figures~\ref{fig:Pi}(a-c) compare the LB and DSMC results for $\Pi$ with respect to 
$T_w$ for $1\ {\rm K} \le T_w \le 3000\ {\rm K}$, at $\delta = 10$, $1$ and $0.1$.
It can be seen that $\Pi$ presents only slight variations with respect to $T_w$ at fixed 
values of $\delta$ and the LB results generally follow the same trend as the 
DSMC results. The general shape of these variations are similar to those of the 
associated viscosity index $\omega$, shown in Fig.~\ref{fig:interp}(b).
The agreement between the LB and DSMC results deteriorates as
$\delta$ is decreased, however the relative error $\Pi_{\rm LB} / \Pi_{\rm DSMC} - 1$
always remains below $2.5\%$, being largest surprisingly at $\delta \simeq 1$.

The simulation results presented in this section for $\delta = 1$ and $10$ 
were obtained using the models ${\rm HHLB}(6;7) \times {\rm HLB}(6;7)$.
As remarked in Refs.~\cite{ambrus16jcp,ambrus18pre}, higher 
orders $Q_x$ of the half-range Gauss-Hermite quadrature are required 
when $\delta \lesssim 1$ in order to get accurate solutions of the 
relaxation time model equation. For this reason, the results pre\-sented 
for $\delta = 0.1$ are obtained using the ${\rm HHLB}(10;16) \times {\rm HLB}(6;7)$ 
model. The values obtained for $\Pi$ \eqref{eq:Pi} using 
these models were within $0.1\%$ of those obtained with the reference 
model ${\rm HHLB}(10; 50) \times {\rm HLB}(6;7)$ on a grid employing
$S = 48$ points. Note that also the values of $\Pi$, obtained with 
the DSMC method, have an error below $0.1\%$ \cite{sharipov18}.

\vspace{-5pt}

\section{CONCLUSION}

\vspace{-2pt}

In this paper, we presented a comparison of lattice Boltzmann (LB) and 
direct simulation Monte Carlo (DSMC) results in the context of Couette flow 
between parallel plates for ${}^3{\rm He}$ and ${}^4{\rm He}$ gases at 
temperatures between $1\ {\rm K}$ and $3000\ {\rm K}$. 
The LB implementation employs the half-range and full-range 
Gauss-Hermite quadratures on the $x$ and $y$ axes, respectively. 
The order $Q_y = 7$ of the full-range Gauss-Hermite quadrature is 
sufficient to obtain accurate results for all tested values of 
the rarefaction parameter $\delta$. The quadrature order on the 
horizontal axis is taken to be $Q_x = 7$ when $\delta = 1$ and 
$\delta = 10$, while at $\delta = 0.1$, it is increased to $Q_x = 16$. 
The total number of velocities is $2Q_xQ_y = 98$ for 
$\delta = 1$ and $\delta = 10$, while at $\delta = 0.1$, 
$224$ velocities are required in order to maintain good accuracy. 
The agreement obtained with the DSMC results 
is very good and the relative error in the 
off-diagonal component $T_{xy}$ of the pressure tensor 
remains under $2.5\%$, even at $\delta = 0.1$. The use of 
the fifth order WENO scheme and of a third order TVD Runge-Kutta 
algorithm allows accurate results to be obtained using only $16$ 
grid nodes on the right half of the channel, which are 
appropriately stretched towards the diffuse reflecting wall.
While the discussion in this paper is limited to 
the Couette flow between parallel plates, we note that the 
LB models based on half-range quadratures can be employed 
also for flows in non-rectangular (curved) domains, 
e.g., when coupled with the vielbein approach, as discussed in 
Ref.~\cite{busuioc19}.

\vspace{-6pt}

\section{ACKNOWLEDGMENTS}

\vspace{-2pt}

V.E.A.~was supported by a grant of the Romanian Ministry of Research and Innovation,
CCCDI-UEFISCDI, project number PN-III-P1-1.2-PCCDI-2017-0371, within PNCDI III.
F.Sh. is supported by CNPq (Brazil), grant 303697/2014-8.

\end{document}